\documentclass[]{emulateapj}
  \usepackage{natbib}
\usepackage{amsfonts}
\usepackage{amssymb}
\usepackage{amsmath,amsthm}
\usepackage{enumerate}
\usepackage{amsmath, amssymb, mathrsfs}
\bibpunct{(}{)}{;}{a}{}{,}

\slugcomment{Accepted for publication in  Astrophysical Journal Letters}
\shorttitle{Phosphorus abundances in Metal-Poor Stars}
\shortauthors{Jacobson et al.}

\begin{document}
\title{The chemical evolution of phosphorus\altaffilmark{1}}

\author{
Heather R.\ Jacobson\altaffilmark{2},
Thanawuth Thanathibodee\altaffilmark{2},
Anna Frebel\altaffilmark{2},
Ian U.\ Roederer\altaffilmark{3},
Gabriele Cescutti\altaffilmark{4},
Francesca Matteucci\altaffilmark{5,6,7}
}

\altaffiltext{1}{Based on observations made with the NASA/ESA {\it
    Hubble Space Telescope}, obtained at the Space Telescope Science
  Institute, which is operated by the Association of Universities for
  Research in Astronomy, Inc., under NASA contract NAS 5-26555.  This
  work is supported through program AR-13246.  Other portions of this
  work are based on data gathered with the 6.5m Magellan Telescopes
  located at Las Campanas Observatory, Chile, and the McDonald
  Observatory of the University of Texas at Austin.}

\altaffiltext{2}{Kavli Institute for Astrophysics and Space Research
  and Department of Physics, Massachusetts Institute of Technology, 77
  Massachusetts Avenue, Cambridge, MA 02139, USA; hrj@mit.edu}

\altaffiltext{3}{Department of Astronomy, University of Michigan, 
  1085 S.\ University Ave., Ann Arbor, MI 48109, USA}

\altaffiltext{4}{Leibniz-Institut f\"{u}r Astrophysik Potsdam (AIP),
  An der Sternwarte 16, D-14482 Potsdam, Germany}

\altaffiltext{5}{Dipartimento di Fisica, Sezione di Astronomia,
  Universit\`{a} di Trieste, Via G. B. Tiepolo 11,  I-34100
  Trieste, Italy}

\altaffiltext{6}{INAF Osservatorio Astronomico di Trieste, Via
  G. B. Tiepolo 11, I-34100 Trieste, Italy}

\altaffiltext{7}{INFN, Sezione di Trieste, Via Valerio 2, 
  I-34100 Trieste, Italy}

\begin{abstract}
Phosphorus is one of the few remaining light elements for which little
is known about its nucleosynthetic origin and chemical evolution,
given the lack of optical absorption lines in the spectra of
long-lived FGK-type stars.  We have identified a P~I doublet in the
near-ultraviolet  (2135/2136 \AA) that is measurable in stars of low metallicity.
Using archival {\it Hubble Space Telescope}-STIS spectra, we have measured P abundances
in 13 stars spanning $-3.3 \le \rm[Fe/H] \le -0.2$, and obtained an
upper limit for a star with $\rm[Fe/H]\sim -3.8$.  Combined with the
  only other sample of P abundances in solar-type stars in the literature, which spans a
  range of $-1 \le \rm[Fe/H] \le +0.2$, we compare the stellar data to
  chemical evolution models.  Our results support previous
  indications that massive-star P yields may need to be increased
  by a factor of a few
  to match stellar data at all metallicities.  Our results  also show
  that hypernovae were important contributors to the
  P production in the early universe.  As P is one of the key
  building blocks of life, we also discuss the chemical evolution of the
  important elements to life, C-N-O-P-S, together.
\end{abstract}

\section{Introduction}\label{sec:intro}

The production of chemical elements over cosmic time is mapped out
well by the element abundances of different metal-poor stars.   The
previous decades have seen an accumulation of abundance
data for most of the element groups in the Periodic Table in 
large samples of stars over a wide metallicity range,
including the most metal-poor stars.  
In contrast, stellar phosphorus (P)
abundances have been striking in their absence.
The paucity of abundance information for phosphorus is due to the
scarcity of P absorption features in the spectra of FGK-type stars.
The only study to trace the P abundances of stars over a range of
metallicity (and thus ``time'' in the chemical evolution of the
universe) is that of \citet{caffau2011}, who analyzed weak P features in the
near-IR.  However, these features become too weak to be measurable in
stars with $\rm [Fe/H] \lesssim -1$, which means little is known of
the chemical evolution of P in the universe at the earliest times.
(The P abundances of some metal-poor damped Ly$\alpha$ (DLA) systems give
a hint, but the sample size is very small; see later sections.)

The results of \citet{caffau2011} show that  
$\rm [P/Fe]$ increases with decreasing $\rm [Fe/H]$ from roughly solar
metallicity down to $\rm [Fe/H] \sim -1$, with hints that it may
plateau at $\rm [P/Fe] \approx +0.2$ below that.  
Comparison of these results to 
chemical evolution models of phosphorus has indicated that current
massive star yields of P must be increased in order
to match the stellar abundance distribution \citep{cescutti2012} at
$\sim$solar metallicity.
However, firm conclusions could not be made at the time without knowledge of 
the P abundances in more metal-poor stars, which are necessary to
better constrain the nucleosynthetic origin of P and its early
chemical evolution in the universe.

We have identified a P~I doublet in the ultraviolet (UV; 2135/2136
\AA) that remains strong and measurable even in very low metallicity
stars.  
A search of the {\it Hubble Space Telescope (HST)} MAST data archive
for high-resolution STIS spectra of Milky Way stars spanning this wavelength
region uncovered spectra of 13 stars with a metallicity range of $-3.3
\leq \rm[Fe/H] \leq -0.2$, or three orders of magnitude in metallicity. This
covers almost the entire range of cosmic chemical
evolution.  The P~I doublet is present in the spectra of all these
stars, allowing us to measure its abundance in stars with 
$\rm [Fe/H] \lesssim -1$ for the first time.

\begin{figure*}[!ht]
\begin{center}
   \includegraphics[clip=true,width=14cm]{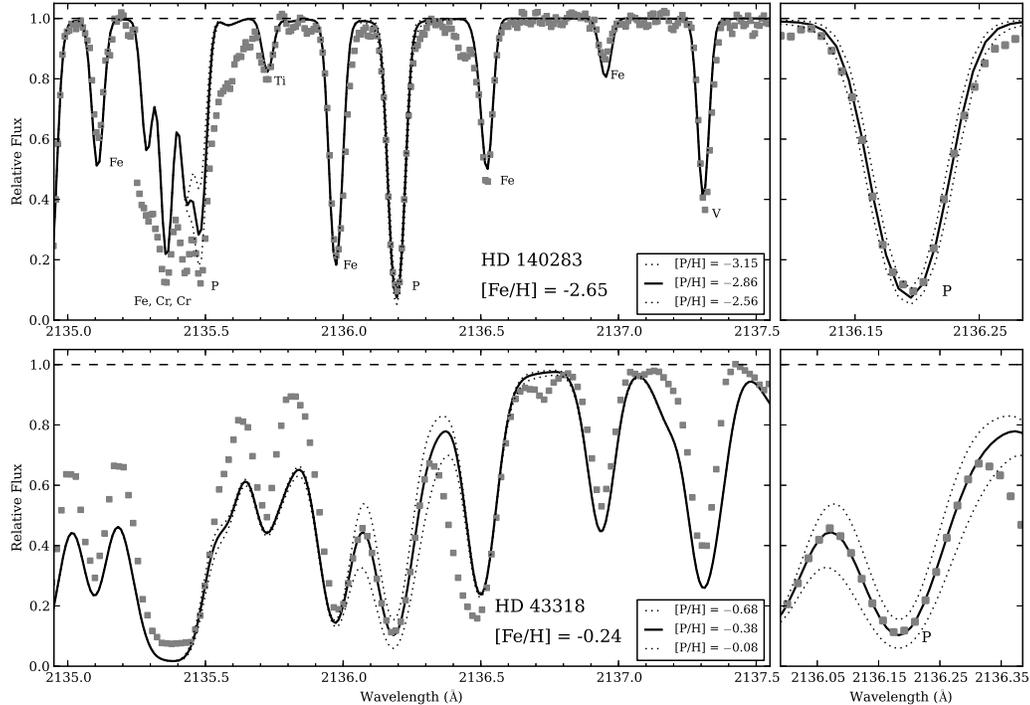} 
     \figcaption{ \label{f_exspec}
       The P~I 2136 \AA\ region in the STIS spectra of HD~140283
       (spectral resolution $R=(\lambda/\Delta\lambda)$ = 110,000) and
       HD~43318 (R = 30,000).  The data are represented by gray squares.  
       Absorption features are identified in the top panels.
       Synthetic spectra of different P abundances are
       plotted over each spectrum, with the best fit indicated by a
       bold line.  A zoom-in on the 2136 \AA\ line is shown in the
       right panels.  Note that the P abundances shown here include
       the empirical 0.23 dex correction described in the text.
}
 \end{center}
\end{figure*}

\section{The P~I 2135/2136 \AA\ doublet}
The P~I doublet at 2135.47 \AA\ and 2136.18 \AA\ (E.P. = 1.41 eV, \citealt{NIST_ASD})
connects the lower
3{\it s}$^{2}$3{\it p}$^{3}$  $^{2}${\it D}$^{o}$ level to the upper
3{\it s}$^{2}$3{\it p}$^{2}$($^{3}${\it P})4{\it s} $^{2}${\it P}
level.  The 2135 \AA\ feature is
often blended with a nearby Cr~II line, but the stronger 2136 \AA\ is
free of blends, even in the solar metallicity stars.  Though this line
becomes saturated in the higher metallicity stars in our sample, 
 comparisons to synthetic spectra show that it remains sensitive enough to 
abundance to be a useful
abundance indicator (see the next section and Figure~\ref{f_exspec}). 
Therefore, we have used this feature to measure
P abundances in our entire stellar sample.

We have adopted the NIST log~{\it gf} value for this line, $-$0.11, for which
they report an accuracy of 25\%.  The 2136 \AA\ feature can exhibit
strong damping wings in the more metal-rich stars.  
We have calculated the van der Waals damping constant,
  $\Gamma_{6}$, for this line using the collisional cross-section
  information from the work of \citet{ABO-1995}.

To our knowledge, no non-LTE calculations have been performed for P~I
features.  A comparison of the abundances of Fe~I and Fe~II lines in the
near-UV (NUV) spectra of six stars in our sample shows abundance differences of 
0.07 dex in the mean, indicating that relatively small non-LTE effects 
may be present, at least for Fe~I.  If non-LTE effects of 
similar magnitude are present for P~I, they have little impact on our
results and are a small component of our total error budget 
(see the next section).  Non-LTE
calculations for P~I are desirable for future studies.

\section{Analysis}\label{analysis}

This work has made use of both NUV and optical
high-resolution spectra available in a number of archive
facilities, as well as spectra obtained by us on the Magellan 6.5m
Clay and 2.7m Harlan J. Smith Telescopes at Las Campanas Observatory and
McDonald Observatory, respectively.  The NUV spectra are in two
spectral resolution groups: $R=(\lambda/\Delta\lambda)$ = 110,000 and
30,000.  The signal-to-noise ratio (S/N) of each spectrum ranges from
10 to 100, with a mean value of 64.  
We refer the reader to \citet{iur-p} for more details about the data.

Although stellar parameters exist in the literature for all stars in
our sample, we performed our own analysis to ensure homogeneity.
We determined stellar parameters and element
abundances from the optical spectra following the method described in
\citet{teff_calib}, and using software developed by A.~Casey \citep{smh}.  
This method employs standard classical
spectroscopic techniques for determining parameters, but applies an
empirical correction to stellar effective temperature to place it on a
scale consistent with photometric temperature determinations.  Our
analysis uses the model atmosphere grid of \citet{castelli_kurucz} and
the LTE analysis code MOOG \citep{moog} which accounts for 
Rayleigh scattering as described in \citet{sobeck11}.  Throughout, 
we adopt the solar abundances of \citet{asplund09}.

\begin{figure*}[!t]
\begin{center}
   \includegraphics[clip=true,width=14cm]{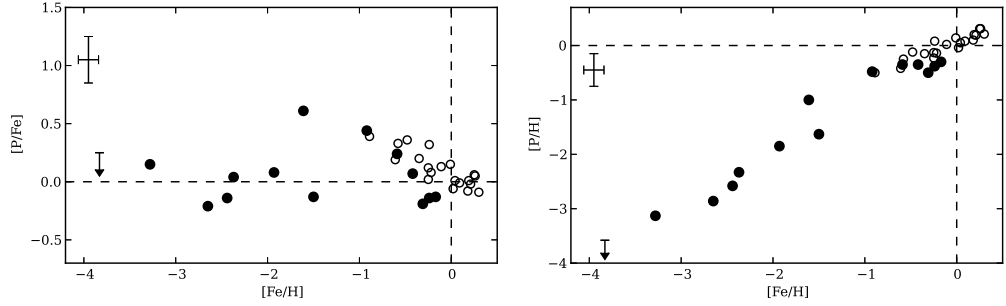} 
     \figcaption{ \label{f_starsp}
       $\rm [P/Fe]$ and $\rm[P/H]$ versus $\rm[Fe/H]$ for the current
       study (filled circles and upper limit symbol) along with the
       sample of \citet{caffau2011} (open circles).  A representative errorbar for an
       individual star is given in the top left corner of each panel.
}
 \end{center}
\end{figure*}

Abundances for 13
 elements (Na, Mg, Al, Si, Ca, Sc, Ti, Cr, Mn, Fe, Co, Ni, and Zn) 
with optical absorption lines were
determined from these spectra, using line lists in
\citet{roederer10} and \citet{ramirez2013}.
Together with the stellar
parameters, these abundances were used as inputs to create detailed synthetic spectra
of the 
region around the P~I doublet for each star.  The P abundance in each synthetic
spectrum was varied until it best matched the P~I 2136 \AA\ feature in the
stellar spectrum.  The best match was determined by inspection
of the residuals of the difference between the observed and synthetic
spectra.  Figure~\ref{f_exspec} illustrates this for two stars in our
sample.  Several other species (predominantly Fe-peak elements) are
also detected in this spectral region, which motivated our use of
the abundances determined from the optical spectra to generate the
synthetic spectra of this region.  As indicated in
Figure~\ref{f_exspec}, the overall agreement between the synthetic and
observed spectra was relatively poorer for the more metal-rich stars.
However, the quality of the fit to these other lines did not affect
measurement of the P feature, and they are largely shown for illustration.

\begin{deluxetable}{lcccrcc}
\tablewidth{0pt}
\tabletypesize{\scriptsize}
\tablecaption{\label{t_pabund} Stellar P and Fe Abundances}
\tablehead{
\colhead{Star} & \colhead{$\rm[Fe/H]$\tablenotemark{a}} &
\colhead{$\sigma$} &
\colhead{$\rm [P/H]$} &
\colhead{$\sigma$} & \colhead{$\rm[P/Fe]$} & 
\colhead{$\sigma$}
}
\startdata
BD~$+$44~493 & $-$3.83\tablenotemark{b} & 0.19 & $< -$3.58 & \nodata & $<$ $+$0.25 & \nodata \\
G~64$-$12    & $-$3.28 & 0.09 & $-$3.13 & 0.34 & $+$0.15 & 0.26 \\
HD~2454      & $-$0.42 & 0.12 & $-$0.35 & 0.27 & $+$0.07 & 0.21 \\
HD~16220     & $-$0.31 & 0.11 & $-$0.50 & 0.26 & $-$0.19 & 0.20 \\
HD~43318     & $-$0.24 & 0.11 & $-$0.38 & 0.26 & $-$0.14 & 0.20 \\
HD~76932     & $-$0.92 & 0.12 & $-$0.48 & 0.29 & $+$0.44 & 0.20 \\
HD~94028     & $-$1.61 & 0.13 & $-$1.00 & 0.29 & $+$0.61 & 0.21 \\
HD~107113    & $-$0.59 & 0.11 & $-$0.35 & 0.26 & $+$0.24 & 0.20 \\
HD~108317    & $-$2.44 & 0.11 & $-$2.58 & 0.35 & $-$0.14 & 0.20 \\
HD~128279    & $-$2.37 & 0.12 & $-$2.33 & 0.36 & $+$0.04 & 0.22 \\
HD~140283    & $-$2.65 & 0.08 & $-$2.86 & 0.30 & $-$0.21 & 0.19 \\
HD~155646    & $-$0.17 & 0.07 & $-$0.30 & 0.25 & $-$0.14 & 0.19 \\
HD~160617    & $-$1.93 & 0.08 & $-$1.85 & 0.27 & $+$0.07 & 0.17 \\
HD~211998    & $-$1.50 & 0.11 & $-$1.63 & 0.33 & $-$0.13 & 0.21 \\
\enddata
\tablenotetext{a}{These values determined from the optical spectra.}
\tablenotetext{b}{From \citet{ito2013}}
\end{deluxetable}

 As previous studies of NUV stellar spectra have uncovered
systematic offsets in element abundances compared to those found
from optical lines (e.g., \citealt{roederer2012}), we also determined
NUV Fe~I and Fe~II abundances for a subset of our sample.  
In order to avoid potential non-LTE
effects on the NUV Fe~I abundances, we restricted the optical-NUV 
abundance comparison to Fe~II lines that are thought to be
unaffected.  Based on six stars in which a
minimum of five NUV Fe~II lines could be measured, the mean
(optical$-$NUV)
offset in Fe~II abundances was found to be $+$0.23$\pm$0.02 ($\sigma$=0.09).
  The difference between this offset and that found by
\citet{roederer2012} in a similar exercise is 0.14 dex (for Fe~II).  
To test that this correction is appropriate for the NUV P
abundances, we adjusted the NUV continuous opacity to force agreement
between NUV and optical Fe~I and Fe~II abundances and remeasured P
abundances.  [P/Fe] ratios agreed within an average of 0.04 dex with
those found by applying the 0.23 dex offset, indicating that this empirical
correction is sufficient.  

Table~\ref{t_pabund} gives the P abundances of our stellar sample,
with this empirical NUV-optical correction applied.  The errors in the
$\rm [P/H]$ and $\rm [P/Fe]$ abundances take into account all
relevant statistical and systematic uncertainties due to errors in
atomic data, stellar parameters, continuum fitting, and the empirical
correction described above.  Such uncertainties include 0.12 dex for the
log~{\it gf}, 0.05 dex for the damping constant, and a 0.14
dex zero-point uncertainty in the NUV-optical abundance correction.
 P abundance sensitivities due to uncertainties
in atmospheric parameters  were assessed for each star and the
  means 
are: 0.07 dex for $\Delta$T$_{\rm eff}$
= 100 K; 0.04 for $\Delta$log~{\it g} = 0.3 dex; 0.06 dex for
$\Delta$v$_{t}$ = 0.2 km s$^{-1}$.  Sensitivity due to errors in
continuum placement are typically no more than 0.07 dex, save for
G~64$-$12, which has S/N$\sim$10 and the weakest P feature. For this
star, uncertainty in the continuum normalization can vary the P
abundance by as much as 0.15 dex.

Finally, we supplement our sample with BD~$+$44~943, the STIS spectrum
of which was recently presented by \citet{placco2014}.  For the
analysis, we have adopted the stellar parameters and metallicity of
\citet{ito2013}.  Only an upper limit to its P abundance could be
determined from the P~I 2553 \AA\ line.

\section{Results and Discussion}\label{results}
\subsection{The Nucleosynthetic Origin(s) of P}

Figure~\ref{f_starsp} shows the P abundance results of the current
sample (filled symbols and upper limit) and those of
\citet{caffau2011} (open circles).  Below $\rm[Fe/H]$ $\sim -2$, 
the $\rm[P/Fe]$ $\sim$ $+$0.0 trend 
(albeit with a large scatter)
depicts the production of P from massive stars. From $\rm[Fe/H]$ $\sim$ $-$1.5,
$\rm[P/Fe]$ then starts to increase with increasing metallicity before
returning to down to $\rm[Fe/H]$ $\sim$0 near solar metallicity.
This behavior is very similar to that predicted by the chemical
evolution models of \citet{kobayashi2006} and \citet{cescutti2012}.
Perhaps more
striking is the distribution of $\rm[P/H]$ versus $\rm[Fe/H]$ (right panel),
which clearly shows a smooth and very well defined buildup of over
$\sim$4 orders of magnitude in P with chemical time (increasing
$\rm[Fe/H]$).   There is reasonable agreement ($\sim$0.25 dex, 
$\approx$2$\sigma$)
with the \citet{caffau2011}
sample P abundances for stars of similar metallicity, with
our values being lower.  This comes after applying the 0.23 dex offset
to our NUV abundance scale.  The UV-IR agreement thus depends on this
offset.  However, we have
  assumed that no additional offset must be applied to place our
  abundances on the same scale as their near-IR abundances.

\begin{figure}[!ht]
\begin{center}
   \includegraphics[clip=true,width=8cm]{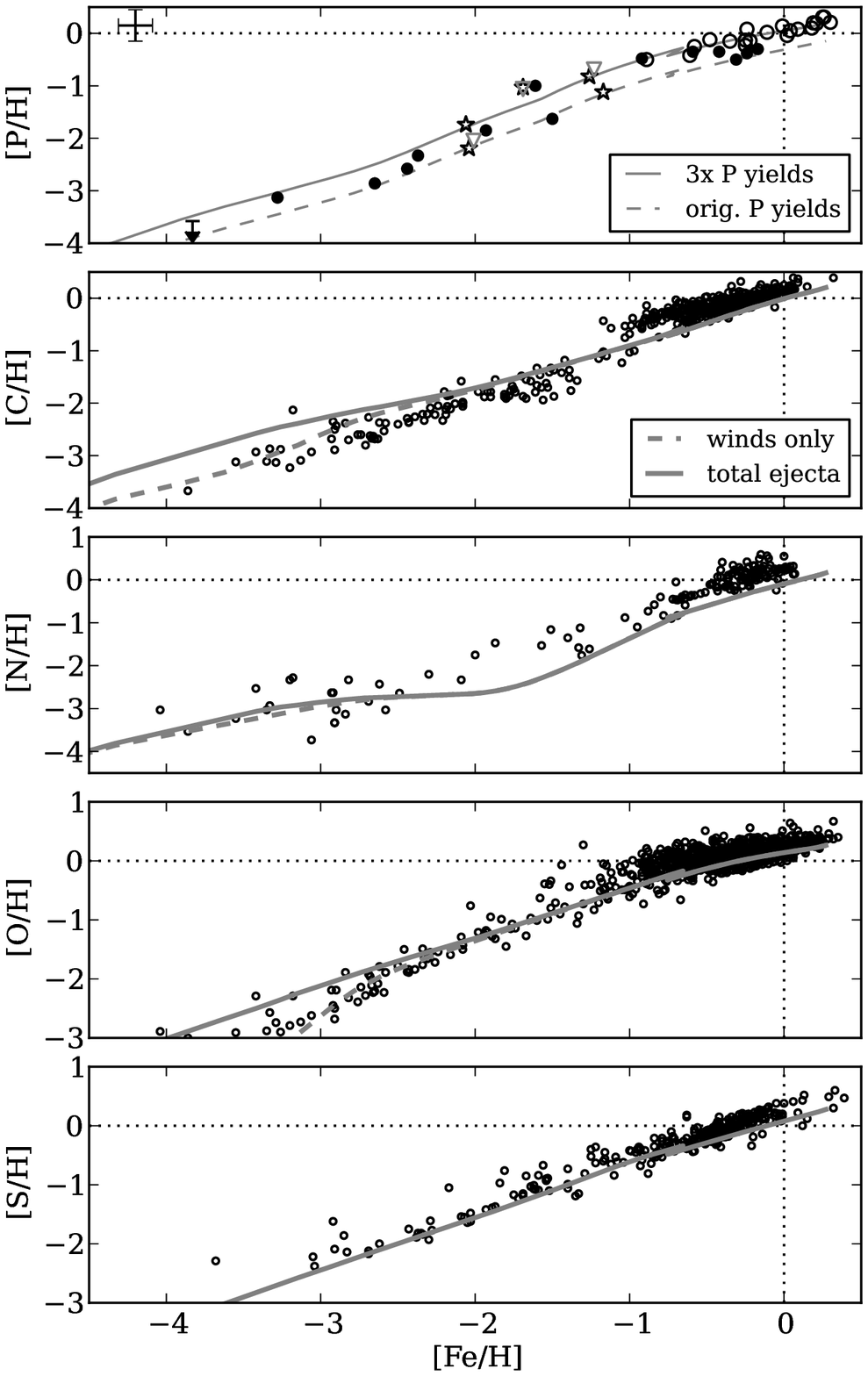} 
     \figcaption{ \label{gce_vh}
      {\it Top panel}: Stellar [P/H] versus [Fe/H] 
      as in Figure~\ref{f_starsp}, with the
      addition of DLA measures (stars) and DLA upper limits (triangles).
      Also shown are the chemical evolution model of
      \citet{cescutti2012}, with massive star P yields from
      \citet{kobayashi2006} (dashed line) and the yields increased by
      a factor of three (solid line).  These correspond to ``Model 6''
      and ``Model 8'' in \citet{cescutti2012}, respectively.
      {\it Remaining panels}: The same for elements C, N, O and S,
      with stellar abundances taken from the literature. 
      Also in each panel is the prediction for the
      chemical evolution model for each element.  For C, N, and O, two
      models are shown, the only difference being the
      treatment of yields from massive stars with Z$< 10^{-5}$
      (spinstars).  The solid lines indicate total C,N,O ejecta (winds
      plus SNe), while the dashed lines represent element production
      in spinstar winds only.  See \citet{cescutti2010} for details. 
      References for the literature stellar samples
      shown here are: \citet{
      israelian2004,ramirez2013,nissen2004,caffau2005,spite2005,lai2008,
      takada-hidai2002,carbon1987,reddy2003,reddy2006,fabbian2009,
      shi2002,nissen2007,takeda2011,spite2011,ryde2004,hansen2011,
      israelian&rebolo2001}.
}
 \end{center}
\end{figure}

\begin{figure}[!t]
\begin{center}
   \includegraphics[clip=true,width=8cm]{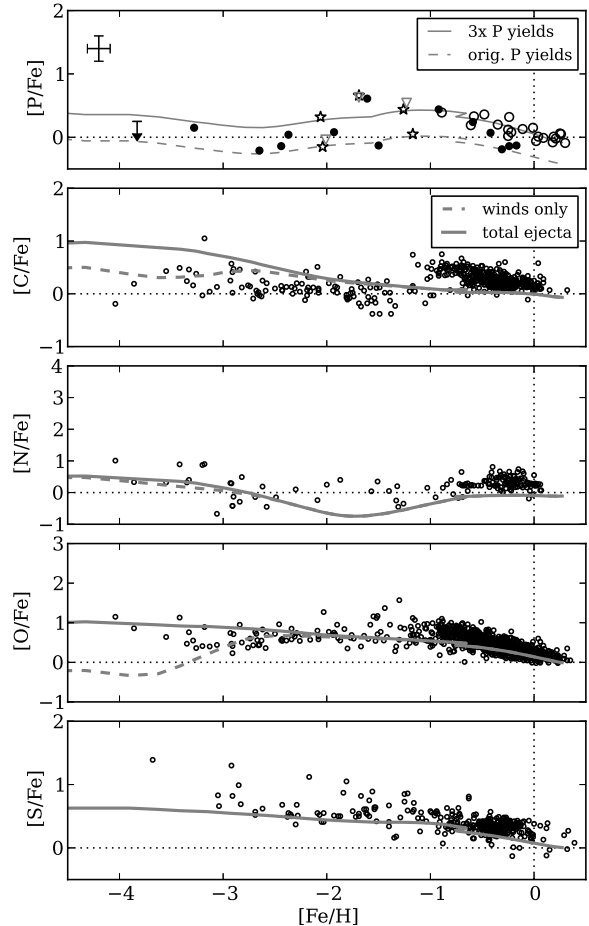} 
     \figcaption{ \label{gce_vfe}
      [X/Fe] versus [Fe/H] for elements P, C, N, O, and S.  Symbols
      and models same as in Figure~\ref{gce_vh}.
}
 \end{center}
\end{figure}

The top panels of Figures~\ref{gce_vh} and \ref{gce_vfe} show the stellar P abundance distributions again,
this time with the results of chemical evolution models from
\citet{cescutti2012}.  The lines in each
panel present chemical evolution model results for the Milky Way,
based on a two-infall model \citep{chiappini1997}, and adopting P
yields for massive stars from \citet{kobayashi2006}.
Also shown here are the P abundances of four DLA systems, 
with upper limits for three more 
(\citealt{molaro1998,molaro2001,outram1999,levshakov2002,lopez2002,fenner2004,rix2007}; 
       R.~Cooke, 2014, private communication).  
The agreement of the independently determined DLA P
abundances with those of the stars is remarkable.

As mentioned in the Introduction, \citet{cescutti2012}
indicated that current supernovae (SNe) P yields must be enhanced by a factor of
$\sim$3 to match the then-available abundance data \citep{caffau2011},
 and especially to obtain [P/Fe]=0 at solar metallicity.  
As can been seen in Figures~\ref{gce_vh} and
\ref{gce_vfe}, most of the stars in our sample indicate that while
some enhancement to the yields is needed, the exact factor is
difficult to derive, especially at low $\rm[Fe/H]$ and given the sizable
(~0.3 dex) error bar on our P abundances. In agreement with
\citet{cescutti2012}, factors of 1-3 are likely and we rule out a
decreased yield.
 
However, even though our results cannot definitely say how much the P
yields must be increased, they do 
 allow us to better
constrain the nucleosynthetic origin for P.
In their work, \citet{cescutti2012} found that two versions of their
chemical evolution model, both with P yields enhanced by $\sim$3, 
fit the \citet{caffau2011} data equally well:
the main difference between the two models, ``Model 7'' and ``Model 8,''
was the inclusion of hypernovae.  
``Model 7'' assumed an energy of 10$^{51}$ erg for all
massive stars, while ``Model 8'' adopted 10$^{52}$ erg for stars with
masses $> 20 M_{\odot}$ (hypernovae).
While exhibiting similar P
abundances at $\sim$solar metallicity, these models showed very
different behavior at low $\rm [Fe/H]$, with 
``Model 7'' having $\rm [P/Fe] > 0.5$ at all $\rm [Fe/H]$ below $-1$.
Our results, most of which have  $\rm [P/Fe] \sim$ 0, 
clearly support ``Model
8'' and indicate that hypernovae were important
sources of P in the early universe.  This is consistent with results
of studies fitting other element abundance patterns of $\rm [Fe/H] \sim -3$
stars, which also found that regular core-collapse SNe were
insufficient to reproduce the data and required contributions from
more energetic events (e.g., \citealt{cayrel2004, tominaga07_b}).

\subsection{The Chemical Evolution of C-N-O-P-S, the Building Blocks
  of Life}
Phosphorus serves a crucial role
in forming and sustaining life: P ions form the backbone of 
deoxyribonucleic acid (DNA) and
are primary constituents in energy exchange in cells, in the form of
adenosine triphosphate (ATP) and adenosine diphosphate (ADP).
Recent research indicates that P, in the form of
polyphosphates, may have played a key role in chaperoning protein
folding when life first formed on Earth billions of years ago
\citep{gray_polyphosphate}.
Now that we
have a clearer view of its formation history, we can review the
chemical evolution of all the building blocks of life, C-N-O-P-S, together.
The remaining panels of 
Figures~\ref{gce_vh} and \ref{gce_vfe} show the cosmic evolution C, N, O and S, based on
literature data, and compared to the same chemical evolution model as
for P.  
For C-N-O, we follow the prescription of \citet{cescutti2010}: the
dashed and solid lines in those panels indicate different
spin star (massive stars with Z$< 10^{-5}$) yields (dashed: from winds
only; solid: ejecta from winds and SNe).  We also limit the stellar
samples for these elements to dwarfs and ``unmixed'' giant stars, as
in \citet{cescutti2010}.
Finally, for S we use the
hypernovae yields from \citet{kobayashi2006}.
Contrary to yield predictions for P until now, the
yields of C, N, O and S are relatively well understood, having been
verified by stellar abundances from many previous studies
\citep{cescutti2010,spite2005,caffau2005}. 

We present Figures~\ref{gce_vh} and \ref{gce_vfe} to highlight a couple of points.  First, they
illustrate the accumulation of these elements in the
universe over cosmic time (right panels), as well as highlight the
importance of massive stars as the main sources of C-N-O-P-S production
at early epochs in the universe (the generally enhanced $\rm[X/Fe]$
values at low metallicities that then decline with the advent of
Type~Ia SNe that produced much Fe, but relatively smaller amounts of the
lighter elements).  Second, they also emphasize the relative degrees to
which we understand the origin and production of these elements:
chemical evolution models with current yields match the observational
data well for C, O, and S, and now P.  This is less true for N, for which
it is difficult to measure primordial abundances in metal-poor stars, and which (along with C) is subject to variations
due to stellar rotation and stellar evolution.

\section{Conclusions}\label{conc}
We have identified a P~I doublet in the NUV that allows for the
measure of P abundances in stars with $\rm [Fe/H] \lesssim -1$ for the
first time.  Using archival high-resolution {\it HST}-STIS spectra of
fourteen stars, we have presented P abundances (and an interesting
upper limit) for stars in the range $-3.8 \le \rm[Fe/H] \le
-0.2$, which spans nearly the entire chemical evolution history of the
universe.  Comparison to chemical evolution models utilizing current
core-collapse SNe P
yields  allows us to test previous indications that P yields must
  be increased by a factor of three to match stellar abundances at all
  metallicities.  Our results indicate that a factor of three may be too
  large to match the observations at low [Fe/H], but that at least some
  enhancement (a factor of $\sim$1.5) is likely necessary.  
Independent of the yield discussion, 
our results clearly imply that hypernovae
were important sources of P production in the early universe, in line
with other results.

Finally, our findings highlight the interdependence of observations of
stars in our galaxy with chemical evolution modeling and nuclear
physics: observations are needed to constrain and revise theoretical
models and nucleosynthesis calculations. Our results also emphasize
the importance of finding bright (V$<$10) very metal-poor stars
observable at high spectral resolution in the UV 
for work such as this.  The study of P and other elements with
features in the UV is entirely dependent on the existence of samples
bright enough for such observations, highlighting the necessity of
dedicated surveys to find these objects \citep{sc14}.

\acknowledgements{
 We thank the referee for comments that improved this manuscript, 
and for directing us to an improved treatment of line damping in this analysis.
R.~Cooke is thanked for providing an unpublished DLA phosphorus measurement.
E.~Toller is acknowledged for her assistance with early testing of the
NUV data analysis.  We also gratefully acknowledge all the
investigators who obtained the observations used in this research, as
well as the institutions that maintain them in public archives.
Based on observations made with the {\it Hubble Space Telescope}
through programs GO-7348, GO-7433, GO-8197, GO-9048, GO-9049, GO-9455, GO-9804, GO-12268, GO-12554, and GO-12976.
Support for program 13246 was provided by NASA through a grant from
the Space Telescope Science Institute, which is operated by the
Association of Universities for Research in Astronomy, Inc., under
NASA contract NAS 5-26555.  Portions of this work are based on data
obtained from the European Southern Observatory (ESO) Science Archive
Facility (programs 065.L-0507(A), 067.D-0439(A), 072.B-0179(A),
074.C-0364(A), 076.B-0055(A), 266.D-5655(A)), as well as data from the
Keck Observatory Archive (KOA), which is operated by the W.M. Keck
Observatory and the NASA Exoplanet Science Institute (NExScI), under
contract with the National Aeronautics and Space Administration
(programs H2aH, H5aH, H47aH).
This research has made
  use of the SIMBAD database, operated at CDS, Strasbourg, France, 
  NASA's Astrophysics Data System Bibliographic Services, and the NIST
  Atomic Spectra Database.  
  A.F.\ acknowledges support from NSF grant AST-1255160.
  T.T.'s work was supported by the MIT UROP program.}

\textit{Facilities:} \facility{{\it HST} (STIS), Magellan:Clay (MIKE), Smith (Tull)}




\begin{thebibliography}{47}
\expandafter\ifx\csname natexlab\endcsname\relax\def\natexlab#1{#1}\fi

\bibitem[{{Anstee} \& {O'Mara}(1995)}]{ABO-1995}
{Anstee}, S.~D. \& {O'Mara}, B.~J. 1995, \mnras, 276, 859

\bibitem[{{Asplund} {et~al.}(2009){Asplund}, {Grevesse}, {Sauval}, \&
  {Scott}}]{asplund09}
{Asplund}, M., {Grevesse}, N., {Sauval}, A.~J., \& {Scott}, P. 2009, ARA\&A,
  47, 481

\bibitem[{{Caffau} {et~al.}(2005){Caffau}, {Bonifacio}, {Faraggiana}, {Fran{\c
  c}ois}, {Gratton}, \& {Barbieri}}]{caffau2005}
{Caffau}, E., {Bonifacio}, P., {Faraggiana}, R., {Fran{\c c}ois}, P.,
  {Gratton}, R.~G., \& {Barbieri}, M. 2005, \aap, 441, 533

\bibitem[{{Caffau} {et~al.}(2011){Caffau}, {Bonifacio}, {Faraggiana}, \&
  {Steffen}}]{caffau2011}
{Caffau}, E., {Bonifacio}, P., {Faraggiana}, R., \& {Steffen}, M. 2011, \aap,
  532, A98

\bibitem[{{Carbon} {et~al.}(1987){Carbon}, {Barbuy}, {Kraft}, {Friel}, \&
  {Suntzeff}}]{carbon1987}
{Carbon}, D.~F., {Barbuy}, B., {Kraft}, R.~P., {Friel}, E.~D., \& {Suntzeff},
  N.~B. 1987, \pasp, 99, 335

\bibitem[Casey(2014)]{smh} Casey, A.~R.\ 2014, 
arXiv:1405.5968

\bibitem[{{Castelli} \& {Kurucz}(2004)}]{castelli_kurucz}
{Castelli}, F. \& {Kurucz}, R.~L. 2004, arXiv:astro-ph/0405087

\bibitem[{{Cayrel} {et~al.}(2004){Cayrel}, {Depagne}, {Spite}, {Hill}, {Spite},
  {Fran{\c c}ois}, {Plez}, {Beers}, {Primas}, {Andersen}, {Barbuy},
  {Bonifacio}, {Molaro}, \& {Nordstr{\"o}m}}]{cayrel2004}
{Cayrel}, R., {Depagne}, E., {Spite}, M., {Hill}, V., {Spite}, F., {Fran{\c
  c}ois}, P., {Plez}, B., {Beers}, T., {Primas}, F., {Andersen}, J., {Barbuy},
  B., {Bonifacio}, P., {Molaro}, P., \& {Nordstr{\"o}m}, B. 2004, A\&A, 416,
  1117

\bibitem[{{Cescutti} \& {Chiappini}(2010)}]{cescutti2010}
{Cescutti}, G. \& {Chiappini}, C. 2010, \aap, 515, A102

\bibitem[{{Cescutti} {et~al.}(2012){Cescutti}, {Matteucci}, {Caffau}, \&
  {Fran{\c c}ois}}]{cescutti2012}
{Cescutti}, G., {Matteucci}, F., {Caffau}, E., \& {Fran{\c c}ois}, P. 2012,
  \aap, 540, A33

\bibitem[{{Chiappini} {et~al.}(1997){Chiappini}, {Matteucci}, \&
  {Gratton}}]{chiappini1997}
{Chiappini}, C., {Matteucci}, F., \& {Gratton}, R. 1997, \apj, 477, 765

\bibitem[{{Fabbian} {et~al.}(2009){Fabbian}, {Nissen}, {Asplund}, {Pettini}, \&
  {Akerman}}]{fabbian2009}
{Fabbian}, D., {Nissen}, P.~E., {Asplund}, M., {Pettini}, M., \& {Akerman}, C.
  2009, \aap, 500, 1143

\bibitem[{{Fenner} {et~al.}(2004){Fenner}, {Prochaska}, \&
  {Gibson}}]{fenner2004}
{Fenner}, Y., {Prochaska}, J.~X., \& {Gibson}, B.~K. 2004, \apj, 606, 116

\bibitem[{{Frebel} {et~al.}(2013){Frebel}, {Casey}, {Jacobson}, \&
  {Yu}}]{teff_calib}
{Frebel}, A., {Casey}, A.~R., {Jacobson}, H.~R., \& {Yu}, Q. 2013, \apj, 769,
  57

\bibitem[{{Gray} {et~al.}(2014){Gray}, {Wholey}, {Wagner}, {Cremers},
  {Mueller-Schickert}, {Hock}, {Krieger}, {Smith}, {Bender}, {Bardwell}, \&
  {Jakob}}]{gray_polyphosphate}
{Gray}, M.~J., {Wholey}, W.-Y., {Wagner}, N.~O., {Cremers}, C.~M.,
  {Mueller-Schickert}, A., {Hock}, N.~T., {Krieger}, A.~G., {Smith}, E.~M.,
  {Bender}, R.~A., {Bardwell}, J.~C.~A., \& {Jakob}, U. 2014, Molecular Cell,
  53, 689

\bibitem[{{Hansen} {et~al.}(2011){Hansen}, {Nordstr{\"o}m}, {Bonifacio},
  {Spite}, {Andersen}, {Beers}, {Cayrel}, {Spite}, {Molaro}, {Barbuy},
  {Depagne}, {Fran{\c c}ois}, {Hill}, {Plez}, \& {Sivarani}}]{hansen2011}
{Hansen}, C.~J., {Nordstr{\"o}m}, B., {Bonifacio}, P., {Spite}, M., {Andersen},
  J., {Beers}, T.~C., {Cayrel}, R., {Spite}, F., {Molaro}, P., {Barbuy}, B.,
  {Depagne}, E., {Fran{\c c}ois}, P., {Hill}, V., {Plez}, B., \& {Sivarani}, T.
  2011, \aap, 527, A65

\bibitem[{{Israelian} {et~al.}(2004){Israelian}, {Ecuvillon}, {Rebolo},
  {Garc{\'{\i}}a-L{\'o}pez}, {Bonifacio}, \& {Molaro}}]{israelian2004}
{Israelian}, G., {Ecuvillon}, A., {Rebolo}, R., {Garc{\'{\i}}a-L{\'o}pez}, R.,
  {Bonifacio}, P., \& {Molaro}, P. 2004, \aap, 421, 649

\bibitem[{{Israelian} \& {Rebolo}(2001)}]{israelian&rebolo2001}
{Israelian}, G. \& {Rebolo}, R. 2001, \apjl, 557, L43

\bibitem[{{Ito} {et~al.}(2013){Ito}, {Aoki}, {Beers}, {Tominaga}, {Honda}, \&
  {Carollo}}]{ito2013}
{Ito}, H., {Aoki}, W., {Beers}, T.~C., {Tominaga}, N., {Honda}, S., \&
  {Carollo}, D. 2013, \apj, 773, 33

\bibitem[{{Kobayashi} {et~al.}(2006){Kobayashi}, {Umeda}, {Nomoto}, {Tominaga},
  \& {Ohkubo}}]{kobayashi2006}
{Kobayashi}, C., {Umeda}, H., {Nomoto}, K., {Tominaga}, N., \& {Ohkubo}, T.
  2006, \apj, 653, 1145

\bibitem[{Kramida {et~al.}(2014)Kramida, {Yu.~Ralchenko}, Reader, \& {and NIST
  ASD Team}}]{NIST_ASD}
Kramida, A., {Yu.~Ralchenko}, Reader, J., \& {and NIST ASD Team}. 2014, {NIST
  Atomic Spectra Database (ver. 5.2), [Online]. Available:
  {\tt{http://physics.nist.gov/asd}} [2014, October 10]. National Institute of
  Standards and Technology, Gaithersburg, MD.}

\bibitem[{{Lai} {et~al.}(2008){Lai}, {Bolte}, {Johnson}, {Lucatello}, {Heger},
  \& {Woosley}}]{lai2008}
{Lai}, D.~K., {Bolte}, M., {Johnson}, J.~A., {Lucatello}, S., {Heger}, A., \&
  {Woosley}, S.~E. 2008, ApJ, 681, 1524

\bibitem[{{Levshakov} {et~al.}(2002){Levshakov}, {Dessauges-Zavadsky},
  {D'Odorico}, \& {Molaro}}]{levshakov2002}
{Levshakov}, S.~A., {Dessauges-Zavadsky}, M., {D'Odorico}, S., \& {Molaro}, P.
  2002, \apj, 565, 696

\bibitem[{{Lopez} {et~al.}(2002){Lopez}, {Reimers}, {D'Odorico}, \&
  {Prochaska}}]{lopez2002}
{Lopez}, S., {Reimers}, D., {D'Odorico}, S., \& {Prochaska}, J.~X. 2002, \aap,
  385, 778

\bibitem[{{Molaro} {et~al.}(2001){Molaro}, {Levshakov}, {D'Odorico},
  {Bonifacio}, \& {Centuri{\'o}n}}]{molaro2001}
{Molaro}, P., {Levshakov}, S.~A., {D'Odorico}, S., {Bonifacio}, P., \&
  {Centuri{\'o}n}, M. 2001, \apj, 549, 90

\bibitem[{{Molaro} {et~al.}(1998){Molaro}, {Vladilo}, \&
  {Centurion}}]{molaro1998}
{Molaro}, P., {Vladilo}, G., \& {Centurion}, M. 1998, \mnras, 293, L37

\bibitem[{{Nissen} {et~al.}(2007){Nissen}, {Akerman}, {Asplund}, {Fabbian},
  {Kerber}, {Kaufl}, \& {Pettini}}]{nissen2007}
{Nissen}, P.~E., {Akerman}, C., {Asplund}, M., {Fabbian}, D., {Kerber}, F.,
  {Kaufl}, H.~U., \& {Pettini}, M. 2007, \aap, 469, 319

\bibitem[{{Nissen} {et~al.}(2004){Nissen}, {Chen}, {Asplund}, \&
  {Pettini}}]{nissen2004}
{Nissen}, P.~E., {Chen}, Y.~Q., {Asplund}, M., \& {Pettini}, M. 2004, \aap,
  415, 993

\bibitem[{{Outram} {et~al.}(1999){Outram}, {Chaffee}, \&
  {Carswell}}]{outram1999}
{Outram}, P.~J., {Chaffee}, F.~H., \& {Carswell}, R.~F. 1999, \mnras, 310, 289

\bibitem[{{Placco} {et~al.}(2014){Placco}, {Beers}, {Roederer}, {Cowan},
  {Frebel}, {Filler}, {Ivans}, {Lawler}, {Schatz}, {Sneden}, {Sobeck}, {Aoki},
  \& {Smith}}]{placco2014}
{Placco}, V.~M., {Beers}, T.~C., {Roederer}, I.~U., {Cowan}, J.~J., {Frebel},
  A., {Filler}, D., {Ivans}, I.~I., {Lawler}, J.~E., {Schatz}, H., {Sneden},
  C., {Sobeck}, J.~S., {Aoki}, W., \& {Smith}, V.~V. 2014, \apj, 790, 34

\bibitem[{{Ram{\'{\i}}rez} {et~al.}(2013){Ram{\'{\i}}rez}, {Allende Prieto}, \&
  {Lambert}}]{ramirez2013}
{Ram{\'{\i}}rez}, I., {Allende Prieto}, C., \& {Lambert}, D.~L. 2013, \apj,
  764, 78

\bibitem[{{Reddy} {et~al.}(2006){Reddy}, {Lambert}, \& {Allende
  Prieto}}]{reddy2006}
{Reddy}, B.~E., {Lambert}, D.~L., \& {Allende Prieto}, C. 2006, \mnras, 367,
  1329

\bibitem[{{Reddy} {et~al.}(2003){Reddy}, {Tomkin}, {Lambert}, \& {Allende
  Prieto}}]{reddy2003}
{Reddy}, B.~E., {Tomkin}, J., {Lambert}, D.~L., \& {Allende Prieto}, C. 2003,
  \mnras, 340, 304

\bibitem[{{Rix} {et~al.}(2007){Rix}, {Pettini}, {Steidel}, {Reddy},
  {Adelberger}, {Erb}, \& {Shapley}}]{rix2007}
{Rix}, S.~A., {Pettini}, M., {Steidel}, C.~C., {Reddy}, N.~A., {Adelberger},
  K.~L., {Erb}, D.~K., \& {Shapley}, A.~E. 2007, \apj, 670, 15

\bibitem[{Roederer} {et~al.}(2014)]{iur-p} Roederer, I.~U., Jacobson,
  H.~R., Thanathibodee, T. \& Frebel, A. 2014, \apjl, accepted

\bibitem[{{Roederer} {et~al.}(2012){Roederer}, {Lawler}, {Sobeck}, {Beers},
  {Cowan}, {Frebel}, {Ivans}, {Schatz}, {Sneden}, \& {Thompson}}]{roederer2012}
{Roederer}, I.~U., {Lawler}, J.~E., {Sobeck}, J.~S., {Beers}, T.~C., {Cowan},
  J.~J., {Frebel}, A., {Ivans}, I.~I., {Schatz}, H., {Sneden}, C., \&
  {Thompson}, I.~B. 2012, \apjs, 203, 27

\bibitem[{{Roederer} {et~al.}(2010){Roederer}, {Sneden}, {Thompson}, {Preston},
  \& {Shectman}}]{roederer10}
{Roederer}, I.~U., {Sneden}, C., {Thompson}, I.~B., {Preston}, G.~W., \&
  {Shectman}, S.~A. 2010, ApJ, 711, 573

\bibitem[{{Ryde} \& {Lambert}(2004)}]{ryde2004}
{Ryde}, N. \& {Lambert}, D.~L. 2004, \aap, 415, 559

\bibitem[{{Schlaufman} \& {Casey}(2014)}]{sc14}
{Schlaufman}, K.~C. \& {Casey}, A.~R. 2014, arXiv:1409.4775

\bibitem[{{Shi} {et~al.}(2002){Shi}, {Zhao}, \& {Chen}}]{shi2002}
{Shi}, J.~R., {Zhao}, G., \& {Chen}, Y.~Q. 2002, \aap, 381, 982

\bibitem[{{Sneden}(1973)}]{moog}
{Sneden}, C.~A. 1973, PhD thesis, The University of Texas at Austin

\bibitem[{{Sobeck} {et~al.}(2011){Sobeck}, {Kraft}, {Sneden}, {Preston},
  {Cowan}, {Smith}, {Thompson}, {Shectman}, \& {Burley}}]{sobeck11}
{Sobeck}, J.~S., {Kraft}, R.~P., {Sneden}, C., {Preston}, G.~W., {Cowan},
  J.~J., {Smith}, G.~H., {Thompson}, I.~B., {Shectman}, S.~A., \& {Burley},
  G.~S. 2011, AJ, 141, 175

\bibitem[{{Spite} {et~al.}(2011){Spite}, {Caffau}, {Andrievsky}, {Korotin},
  {Depagne}, {Spite}, {Bonifacio}, {Ludwig}, {Cayrel}, {Fran{\c c}ois}, {Hill},
  {Plez}, {Andersen}, {Barbuy}, {Beers}, {Molaro}, {Nordstr{\"o}m}, \&
  {Primas}}]{spite2011}
{Spite}, M., {Caffau}, E., {Andrievsky}, S.~M., {Korotin}, S.~A., {Depagne},
  E., {Spite}, F., {Bonifacio}, P., {Ludwig}, H.-G., {Cayrel}, R., {Fran{\c
  c}ois}, P., {Hill}, V., {Plez}, B., {Andersen}, J., {Barbuy}, B., {Beers},
  T.~C., {Molaro}, P., {Nordstr{\"o}m}, B., \& {Primas}, F. 2011, \aap, 528, A9

\bibitem[{{Spite} {et~al.}(2005){Spite}, {Cayrel}, {Plez}, {Hill}, {Spite},
  {Depagne}, {Fran{\c c}ois}, {Bonifacio}, {Barbuy}, {Beers}, {Andersen},
  {Molaro}, {Nordstr{\"o}m}, \& {Primas}}]{spite2005}
{Spite}, M., {Cayrel}, R., {Plez}, B., {Hill}, V., {Spite}, F., {Depagne}, E.,
  {Fran{\c c}ois}, P., {Bonifacio}, P., {Barbuy}, B., {Beers}, T., {Andersen},
  J., {Molaro}, P., {Nordstr{\"o}m}, B., \& {Primas}, F. 2005, \aap, 430, 655

\bibitem[{{Takada-Hidai} {et~al.}(2002){Takada-Hidai}, {Takeda}, {Sato},
  {Honda}, {Sadakane}, {Kawanomoto}, {Sargent}, {Lu}, \&
  {Barlow}}]{takada-hidai2002}
{Takada-Hidai}, M., {Takeda}, Y., {Sato}, S., {Honda}, S., {Sadakane}, K.,
  {Kawanomoto}, S., {Sargent}, W.~L.~W., {Lu}, L., \& {Barlow}, T.~A. 2002,
  \apj, 573, 614

\bibitem[{{Takeda} \& {Takada-Hidai}(2011)}]{takeda2011}
{Takeda}, Y. \& {Takada-Hidai}, M. 2011, \pasj, 63, 537

\bibitem[{{Tominaga} {et~al.}(2007){Tominaga}, {Umeda}, \&
  {Nomoto}}]{tominaga07_b}
{Tominaga}, N., {Umeda}, H., \& {Nomoto}, K. 2007, ApJ, 660, 516

\end{thebibliography}
\end{document}